\documentclass[showpacs,preprintnumbers,onecolumn,12pt]{revtex4}
\usepackage{amssymb}
\usepackage{amsfonts}
\usepackage{amsmath}
\usepackage{graphicx}
\usepackage{dcolumn}
\usepackage{bm}

\input{tcilatex}

\begin{document}

\title{Homoclinic Signatures of Dynamical Localization}
\author{Ricardo Chac\'{o}n}
\affiliation{Departamento de Electr\'{o}nica e Ingenier\'{\i}a Electromec\'{a}nica,
Escuela de Ingenier\'{\i}as Industriales, Universidad de Extremadura,
Apartado Postal 382, E-06071 Badajoz, Spain}
\date{\today}

\begin{abstract}
It is demonstrated that the oscillations in the width of the momentum
distribution of atoms moving in a phase-modulated standing light field, as a
function of the modulation amplitude $\lambda $, are correlated with the
variation of the chaotic layer width in energy of an underlying \textit{%
effective} pendulum. The maximum effect of dynamical localization and the
nearly perfect delocalization are associated with the maxima and minima,
respectively, of the chaotic layer width. It is also demonstrated that
kinetic energy is conserved as an almost adiabatic invariant at the minima
of the chaotic layer width, and that the system is \textit{accurately}
described by delta-kicked rotors at sufficiently large zeros of the Bessel
functions $J_{0}\left( \lambda \right) $ and $J_{1}\left( \lambda \right) $.
Numerical calculations of kinetic energy and Lyapunov exponents confirm all
the theoretical predictions.
\end{abstract}

\maketitle

% Force line breaks with \\




Nonautonomous Hamiltonian systems exhibit a wide range of novel effects with
different manifestations in the quantum and classical domains [1,2]. A
situation of especial interest occurs when the classical dynamics undergoes
a transition from stability to chaos as a control parameter is varied [3].
In this regard, dynamical localization (DL) is a specifically quantum
phenomenon taking place in systems where the classical dynamics exhibits
chaotic diffusion in momentum space, while the asymptotic quantum momentum
distribution freezes to a steady state although it is initially similar to
its classical counterpart. This phenomenon is attributed to quantum
interferences among the diffusive paths which on average over long times are
completely destructive. These interferences giving rise to chaotic diffusion
are properly described in terms of gradual dephasing of the various Floquet
states that contribute to the initial state. After a characteristic time
(the so-called Ehrenfest time $\tau _{E}\sim \ln \left( S/\hbar \right)
/\Lambda ^{+}$, where $\Lambda ^{+}$ is the sum of positive Lyapunov
exponents (LEs) and $S$ is a typical classical action [4]), the momentum
distribution covers all significantly populated Floquet states as these
become completely dephased, leading to the freezing of the diffusive growth.
This phenomenon has been discussed in model systems such as the kicked
rotator [5] as well as in the center-of-mass motion of a single ion confined
in a Paul trap and interacting with a laser field [6]. While no classical or
semi-classical explanation for DL has as yet been proposed, a formal analogy
with the Anderson localization appearing in disordered systems has been
discussed [7]: DL occurs over the time coordinate with localization in
momentum space, while Anderson localization refers to spatial coordinates.
Additional evidence of the DL phenomenon has been provided by the momentum
transfer from a modulated standing light wave to a sample of ultracold atoms
[8-12]. It was found experimentally [9,11] and numerically [10,11] that the
width of the atomic momentum distribution exhibits oscillations as a
function of the modulation amplitude, while a theoretical analysis [8]
predicted its overall monotonous decrease for sufficiently large modulation
amplitudes (\textit{quantum} regime). Also, a preliminary qualitative
analysis of such oscillatory behavior was previously discussed in terms of
the zeros of Bessel functions [10,11]. However, no theoretical approach has
as yet been proposed to explain in detail the waveform of the aforementioned
oscillations or the amplitude values at which the maximum effect of DL had
been found experimentally and numerically. In this Letter, such fundamental
questions in terms of the strength of \textit{homoclinic instabilities} are
explained by studying the dependence on the modulation amplitude of the
width in energy of the chaotic separatrix layer associated with an
underlying effective perturbed pendulum. Remarkably, homoclinic
instabilities \textit{also} imprint a clear signature in the quantum spectra
of a particle confined in a stadium billiard [13].

\textit{Effective model.}$-$The motion of an atom in a phase-modulated light
field, as created for instance by an oscillating mirror, is described [8] by
the Hamiltonian $\widetilde{H}=\widetilde{p}^{2}/(2M)-V_{0}\cos \left[ 2k%
\widetilde{x}-\lambda \sin \left( \omega t\right) \right] $, where $M,%
\widetilde{x},\widetilde{p},V_{0},k,\lambda ,$ and $\omega $ denote the
mass, position, momentum, height of the periodic potential, dimensionless
modulation depth, wave number, and frequency modulation, respectively.
Switching to scaled dimensionless variables $\tau \equiv \omega t,x\equiv 2k%
\widetilde{x},p\equiv \left[ 2k/\left( M\omega \right) \right] \widetilde{p}$%
, $H\equiv \left[ 4k^{2}/\left( M\omega ^{2}\right) \right] \widetilde{H}$,
one obtains the dimensionless Hamiltonian%
\begin{equation}
H=\frac{p^{2}}{2}-\kappa \cos \left( x-\lambda \sin \tau \right) ,  \tag{1}
\end{equation}%
where $x,p,\tau ,$ and $\kappa $ are the dimensionless coordinate, momentum,
time, and height of the potential, respectively. It is worth noting that the
Hamiltonian (1) also describes the dynamics of a small dissipationless
Josephson junction driven by an external sinusoidal current source [14]. To
discuss an energy-based characterization of the chaotic dynamics for $%
\lambda >0$, one rewrites Eq. (1) as $H=H_{0}+H_{1}$ where $H_{0}\equiv
p^{2}/2-\kappa \cos x$, $H_{1}\equiv \kappa \left[ \cos x-\cos \left(
x-\lambda \sin \tau \right) \right] $. Now one assumes that $H$ may be
regarded as the Hamiltonian of an effective perturbed pendulum, with $H_{1}$
being the Hamiltonian perturbation for $\lambda >0$. The change of energy of
the unperturbed pendulum because of the influence of $H_{1}$ can be
determined from $dH_{0}/d\tau =\kappa p[\sin x-\sin \left( x-\lambda \sin
\tau \right) ]$ and hence the change of kinetic energy: $d\left(
p^{2}/2\right) /d\tau =-\kappa p\sin \left( x-\lambda \sin \tau \right) $.
Since homoclinic dynamics is the origin of the chaotic behavior in
Hamiltonian systems, it is useful to calculate the change of the averaged
energy along the separatrix $x_{0,\pm }\left( \tau \right) =\pm 2\arctan
\left\{ \sinh \left[ \sqrt{\kappa }\left( \tau -\tau _{0}\right) \right]
\right\} $, $p_{0,\pm }\left( \tau \right) =\pm 2\sqrt{\kappa }\func{sech}%
\left[ \sqrt{\kappa }\left( \tau -\tau _{0}\right) \right] $ of the
unperturbed pendulum $H_{0}$: $\left\langle dH_{0}/d\tau \right\rangle
\equiv \int_{-\infty }^{\infty }\left( dH_{0}/d\tau \right) d\tau $, where $%
\tau =\tau _{0}$ gives the centre of the soliton [2]. Expanding the term $%
\sin \left( x-\lambda \sin \tau \right) $ by using the relationship $\exp
\left( i\lambda \sin \tau \right) =\sum_{n}J_{n}\left( \lambda \right) \exp
\left( in\tau \right) $ of the Bessel functions, one straightforwardly finds%
\begin{equation}
\left\langle \frac{dH_{0}}{d\tau }\right\rangle =\left\langle \frac{d\left(
p^{2}/2\right) }{d\tau }\right\rangle \approx d\left( \lambda ,\kappa
\right) \equiv \sum_{n=1}^{\infty }\left[ a_{n}^{2}\left( \lambda ,\kappa
\right) +2b_{n}^{2}\left( \lambda ,\kappa \right) +2a_{n}\left( \lambda
,\kappa \right) b_{n}\left( \lambda ,\kappa \right) \right] ^{1/2},  \tag{2}
\end{equation}%
where $a_{n}\left( \lambda ,\kappa \right) \equiv 16\pi n^{2}J_{2n}\left(
\lambda \right) J_{0}^{-2}\left( \lambda \right) \func{csch}\left( n\pi /%
\sqrt{\kappa \left\vert J_{0}\left( \lambda \right) \right\vert }\right)
/\kappa $, $b_{n}\left( \lambda ,\kappa \right) \equiv 4\pi \left\vert
J_{2n-1}\left( \lambda \right) /J_{0}\left( \lambda \right) \right\vert %
\left[ 1+2\left( 2n-1\right) ^{2}/\left\vert \kappa J_{0}\left( \lambda
\right) \right\vert \right] \func{sech}\left[ \left( 2n-1\right) \pi /(2%
\sqrt{\kappa \left\vert J_{0}\left( \lambda \right) \right\vert })\right] $.
The width function $d\left( \lambda ,\kappa \right) $ provides an estimate
of the width in energy of the chaotic separatrix layer. Figure 1(a) clearly
shows that the behavior of the width function $d\left( \lambda \right) $
correlates with those of the classical and quantum mechanical momentum
distributions obtained in Ref. [10]. In particular, Fig. 1(b) shows that the
maxima of $d\left( \lambda \right) $ approximately correspond to the
amplitude values where a maximum effect of DL appears, while the minima of $%
d\left( \lambda \right) $ correspond to the values where classical and
quantum distributions approximately coincide (i.e., DL disappears). Figure 1
also shows that the function $d\left( \lambda \right) $ presents two kinds
of minima: those exactly occurring at the zeros of $J_{0}\left( \lambda
\right) $, where the estimate of the chaotic layer width becomes null, and
those approximately occurring at the zeros of $J_{1}\left( \lambda \right) $%
, where the chaotic layer exhibits a small but finite width. These two kinds
of minima suggest that the kinetic energy could be conserved as an \textit{%
almost adiabatic invariant} [15] at the zeros $\left\{ \lambda _{0}^{\ast
}\right\} $ of $J_{0}\left( \lambda \right) $ for a \textit{certain }set of
initial conditions, and at the zeros $\left\{ \lambda _{1}^{\ast }\right\} $
of $J_{1}\left( \lambda \right) $ for another set of initial conditions.
Numerical calculations of the maximal LE (see Figs. 2-4) and the kinetic
energy (data not shown) confirm these theoretical predictions. Indeed,
system (1) presents three LEs $\left( \Lambda ^{+},0,\Lambda ^{-}\right) $
which satisfy $\Lambda ^{+}=-$ $\Lambda ^{-}$. When all three LEs vanish,
the motion is quasi-periodic and the kinetic energy is conserved in the
aforementioned sense, as in the examples shown in Figs. 3 and 4. Remarkably,
one finds that the function $d\left( \lambda \right) $ also correlates with $%
\Lambda ^{+}$ and hence with $\tau _{E}^{-1}$ (see Fig. 2). Comparison of
Figs. 3 and 4 clearly indicates that DL strongly depends on the initial
momentum, which should be taken into account in experiments [16].

\textit{Dynamics at the zeros of }$J_{0}\left( \lambda \right) $ \textit{and 
}$J_{1}\left( \lambda \right) $.$-$Using well-known properties of $%
J_{n}\left( \lambda \right) $ and $\delta _{T}$ (the Dirac comb of period $T$%
) [17,2], one straightforwardly obtains that Eq. (1) at $\left\{ \lambda
_{0}^{\ast }\right\} ,\left\{ \lambda _{1}^{\ast }\right\} $ reduces to%
\begin{eqnarray}
\frac{d^{2}z}{d\tau ^{\prime \prime 2}}+K_{0}\delta _{T}\left( \tau ^{\prime
\prime }\right) \sin z &=&O\left( 1/\lambda _{0}^{\ast 2}\right) , 
\TCItag{3} \\
\frac{d^{2}x}{d\tau ^{\prime 2}}+K_{1}\delta _{T}\left( \tau ^{\prime
}\right) \sin x &=&O[J_{0}\left( \lambda _{1}^{\ast }\right) /\lambda
_{1}^{\ast }],  \TCItag{4}
\end{eqnarray}%
respectively, where $\tau ^{\prime \prime }\equiv \sqrt{\kappa }\left( \tau
-\pi /4\right) ,\tau ^{\prime }\equiv \sqrt{\kappa }\left( \tau -\pi
/2\right) ,K_{0}\equiv \pi \sqrt{\kappa }\left\vert J_{1}\left( \lambda
_{0}^{\ast }\right) \right\vert /2,K_{1}\equiv \pi \sqrt{\kappa }\left\vert
J_{0}\left( \lambda _{1}^{\ast }\right) \right\vert ,T\equiv \pi \sqrt{%
\kappa },z\equiv x\pm \pi /2$ with the sign $+\left( -\right) $ when $%
J_{1}\left( \lambda _{0}^{\ast }\right) <\left( >\right) 0$, and where the
shift $x\rightarrow x\pm \pi $ is applied when $J_{0}\left( \lambda
_{1}^{\ast }\right) <0$. One sees that the description by means of the $%
\delta $-kicked rotors (3), (4) is almost perfect in the quantum regime
(i.e., for sufficiently large zeros), and that the shift $z\equiv x\pm \pi /2
$ in their spatial coordinates as well as the different values of the
amplitudes $K_{0},K_{1}$ help to understand why the approximate conservation
of the kinetic energy does not occur for the same set of initial conditions
at both kinds of zeros (cf. Figs. 3 and 4). Therefore, Eqs. (3) and (4) can
be reduced in the quantum regime to the standard maps [2]: $\overset{\_}{p}%
=p+K_{0}\sin z,\overset{\_}{z}=z+T\overset{\_}{p}$; $\overset{\_}{p}%
=p+K_{1}\sin x,\overset{\_}{x}=x+T\overset{\_}{p}$, per period $T$,
respectively. In such a case, the energy of a trajectory randomly fluctuates
inside the chaotic separatrix layer whose width is $\Delta E_{s,0}\left(
\lambda _{0}^{\ast },\kappa \right) \approx 64\pi \kappa ^{-1}\left\vert
J_{1}\left( \lambda _{0}^{\ast }\right) \right\vert ^{-1}\exp \left[ -\pi 
\sqrt{2/\left( \kappa \left\vert J_{1}\left( \lambda _{0}^{\ast }\right)
\right\vert \right) }\right] $, $\Delta E_{s,1}\left( \lambda _{1}^{\ast
},\kappa \right) \approx 32\pi \kappa ^{-1}\left\vert J_{0}\left( \lambda
_{1}^{\ast }\right) \right\vert ^{-1}\exp \left( -\pi /\sqrt{\kappa
\left\vert J_{0}\left( \lambda _{1}^{\ast }\right) \right\vert }\right) $,
respectively. Note that these estimates are in excellent agreement with Eq.
(2). For example, for the experimental value $\kappa =0.36$ [9,11] one has $%
\Delta E_{s,0}\left( 8.6537,0.36\right) \simeq 0.00138,\Delta E_{s,1}\left(
7.0156,0.36\right) \simeq 0.06574$ while $d\left( 8.6537,0.36\right) \simeq
0,d\left( 7.0156,0.36\right) \simeq 0.06769$. In general, $\Delta
E_{s,0}\left( \lambda _{0}^{\ast },0.36\right) ,\Delta E_{s,1}\left( \lambda
_{1}^{\ast },0.36\right) $ are very small widths at sufficiently large
values of $\lambda _{0}^{\ast },\lambda _{1}^{\ast }$, which explains the
approximate conservation of the kinetic energy at such zeros for the
majority of the initial conditions.

\textit{Conclusion.}$-$Now an interpretation of the above results in the
framework of Floquet theory is in order: They indicate that increasing the
strength of chaos implies decreasing $\tau _{E}$, i.e., the time interval in
which the Floquet states significantly contributing to the dynamics become
completely dephased. Thus, the maxima of the chaotic layer width could be
associated with corresponding minima of $\tau _{E}$. This means that when
the width in energy of the chaotic layer reaches a maximum as a function of
the modulation amplitude, the corresponding quantum momentum distribution
covers a minimal set of significantly populated Floquet states, and the
effect of DL is maximally enhanced. This picture could well explain some
previously discussed ways of breaking DL, such as deviation from strict
periodicity in the driving [18]. Indeed, it has been demonstrated that the
suppression of chaos [19] and the destruction of DL [18] by a secondary
irrational frequency are both very sensitive phenomena. It is also coherent
with the destruction of DL due to dissipation [20]: decreasing chaos by
increasing dissipation is a universal mechanism [2]. Thus, the present
theoretical approach to DL suggests that diverse methods for controlling
chaos [19] could also be useful in controlling DL. Finally, \textit{exact}
experimental realization of the standard $\delta $-kicked rotor by using
ultracold atoms interacting with a standing wave of laser light can be
achieved at the aforementioned suitable values of the modulation amplitude.

The author acknowledges P. Bardroff \ and W. P. Schleich for kindly
providing the numerical data depicted in Fig. 1. This work was supported by
the Spanish MCyT and the European Regional Development Fund (FEDER) program
through project FIS2004-02475.

\subsection{Figure Captions}

\bigskip

Figure 1. (a) Comparison between the classical [thick blue (black) line] and
quantum mechanical [red (gray) line] momentum distribution of atoms moving
in a phase-modulated standing light field for $\tau =320$ (cf. Ref.[10]).
Here, $\Delta p\equiv p_{RMS}/(2\hbar k)\equiv \left[ \left\langle
p^{2}\right\rangle -\left\langle p\right\rangle ^{2}\right] ^{1/2}/(2\hbar k)
$ is a normalized RMS momentum width. Also plotted for comparison is the
width of the chaotic layer [Eq. (2), thin black line] vs the amplitude $%
\lambda $. (b) Difference between the classical and quantum mechanical
momentum distributions $\Delta p_{C-Q}$ (green line), width of the chaotic
layer (black line), and Bessel functions $3\left\vert J_{1}\left( \lambda
\right) \right\vert $ (dashed line) and $3\left\vert J_{0}\left( \lambda
\right) \right\vert $ vs the amplitude $\lambda $.

\bigskip

Figure 2. Maximal LE corresponding to $\kappa =0.36$ and nine initial
conditions inside the initial well and their average (thick black line), $%
\Lambda ^{+}$, and the rescaled chaotic layer width [Eq. (2), red (gray)
line] as functions of the amplitude $\lambda $. The null value of $\Lambda
^{+}$ for finite ranges of $\lambda $ around the zeros of $J_{1}\left(
\lambda \right) $ indicates quasi-periodic motion.

\bigskip

Figure 3. Grids of $75\times 75$ points in the $x(0)-\lambda $ plane for $%
\kappa =0.36$ and $p(0)=0$. Blank, yellow, red, and black regions indicate
that the maximal LE, $\Lambda ^{+}$, satisfies $\left\vert \Lambda
^{+}\right\vert <10^{-3},\Lambda ^{+}\in \left[ 10^{-3},0.06\right] ,\Lambda
^{+}\in (0.06,0.12],$ and $\Lambda ^{+}>0.12$, respectively. The zeros of $%
J_{0}\left( \lambda \right) $ are also indicated for comparison with the
regions where the kinetic energy is approximately conserved (blank regions).

Figure 4.  Grids of $75\times 75$ points in the $x(0)-\lambda $ plane for $%
\kappa =0.36$ and $\ p(0)=-1$. Blank, yellow, red, and black regions
indicate that the maximal LE, $\Lambda ^{+}$, satisfies $\left\vert \Lambda
^{+}\right\vert <10^{-3},\Lambda ^{+}\in \left[ 10^{-3},0.06\right] ,\Lambda
^{+}\in (0.06,0.12],$ and $\Lambda ^{+}>0.12$, respectively. The zeros of $%
J_{1}\left( \lambda \right) $ are also indicated for comparison with the
regions where the kinetic energy is approximately conserved (blank regions).

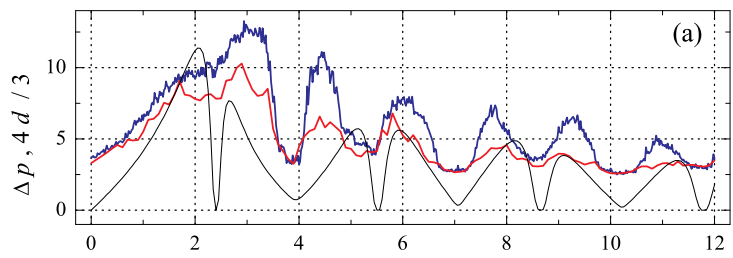
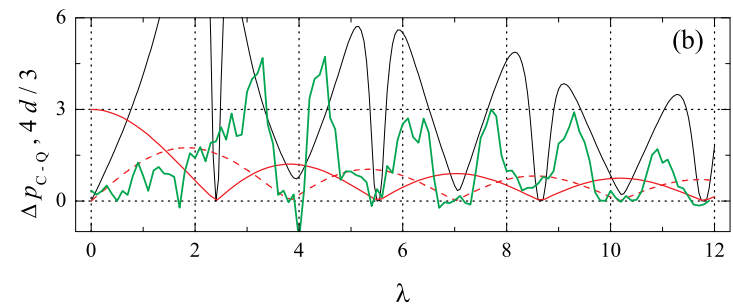
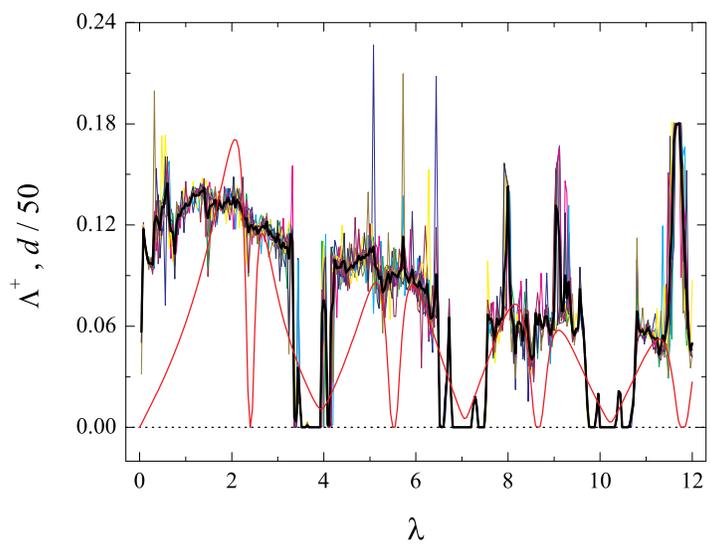
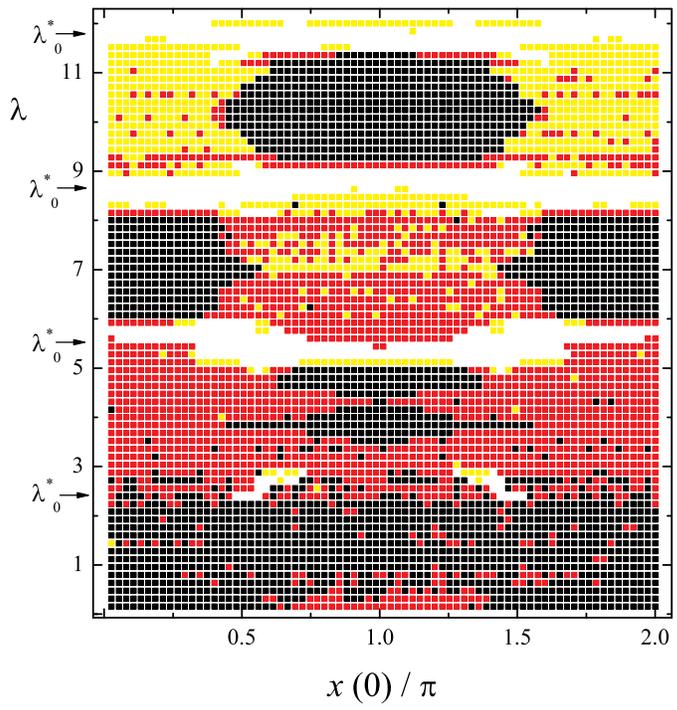
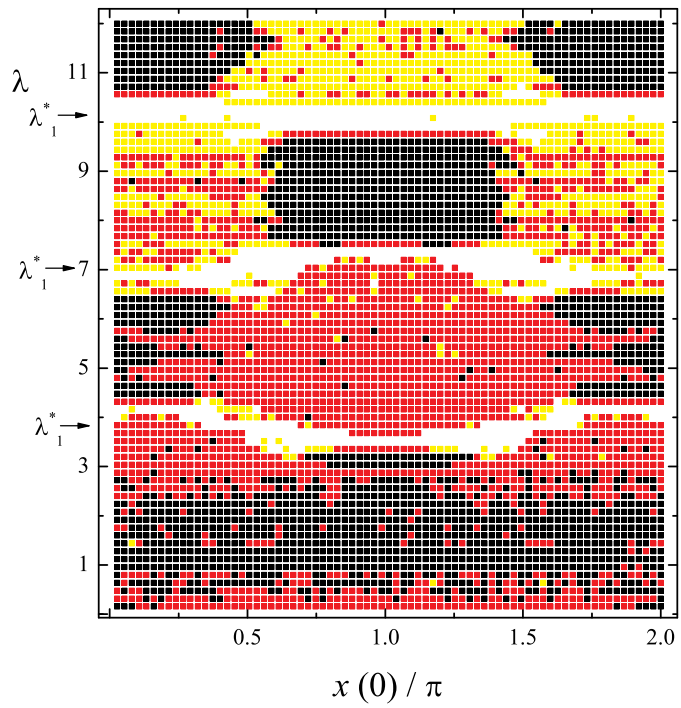
\end{document}